# Near-Infrared Lead Chalcogenide Quantum Dots: Synthesis and Applications in Light Emitting Diodes*

Haochen Liu(刘皓宸)[1], Huaying Zhong(钟华英)[1], Fankai Zheng(郑凡凯)[1], Yue Xie(谢阅)[1], Depeng Li(李德鹏)[1], Dan Wu(吴丹)[2], Ziming Zhou(周子明)[1], Xiao Wei Sun(孙小卫)[1], and Kai Wang(王恺)[1†]

[1] *Guangdong University Key Lab for Advanced Quantum Dot Displays and Lighting, Shenzhen Key Laboratory for Advanced Quantum Dot Displays and Lighting, and Department of Electrical and Electronic Engineering, Southern University of Science and Technology, Xueyuan Blvd. 1088, Shenzhen, 518055, Guangdong, China*
[2] *Academy for Advanced Interdisciplinary Studies, Southern University of Science and Technology, Xueyuan Blvd. 1088, Shenzhen, 518055, Guangdong, China*

This paper reviews recent progress in the synthesis of near-infrared (NIR) lead chalcogenide (PbX; PbX=PbS, PbSe, PbTe) quantum dots (QDs) and their applications in NIR QDs based light emitting diodes (NIR-QLEDs). It summarizes the strategies of how to synthesize high efficiency PbX QDs and how to realize high performance PbX based NIR-QLEDs.

## 1. Introduction

Colloidal semiconductor quantum dots (QDs) are semiconductor materials with a particle size ranging from 2 to 20 nm. They are also called "artificial atoms", as they have a discrete and atom-like energy level structure due to the strong restricted movement of excitons in three dimensions.[1, 2] Different from their bulk counterparts, the optical and electronic properties of QDs can be easily tailored not only by their chemical composition but also by size, shape and the size distribution, which makes QDs attractive in various applications. In particular, because of the colloidal nature of QDs, they can be dispersed in solvents and easily assembled to form solid-state materials using low-cost and large-area solution processing. These advantages of colloidal QDs enable them as promising candidates for mass production including inkjet printing, roll-to-roll deposition, or spraying techniques, which are applicable to the wafer-scale fabrication[3] of a wide range of optoelectronic devices such as solar cells ,[4, 5] light emitting diodes (LEDs),[6, 7] thin film transistors[8, 9] and photodetectors. [10, 11]

* Project supported by National Key Research and Development Program (No.2016YFB0401702, No.2017YFE0120400), National Natural Science Foundation of China (No.61875082, No.61405089), Guangdong University Key Laboratory for Advanced Quantum Dot Displays and Lighting (No. 2017KSYS007), Natural Science Foundation of Guangdong (No.2017B030306010), Guangdong Province's 2018-2019 Key R&D Program: Environmentally Friendly Quantum Dots Luminescent Materials (Project No.: 2019B010924001), Shenzhen Innovation Project (No.JCYJ20160301113356947, No. JSGG20170823160757004), Shenzhen Peacock Team Project (No.KQTD2016030111203005), Shenzhen Key Laboratory for Advanced Quantum Dot Displays and Lighting (No.ZDSYS20170728163632549), and Tianjin New Materials Science and Technology Key Project (No.16ZXCLGX00040).
† Corresponding author. E-mail: wangk@sustech.edu.cn

Compared with the development of visible colloidal QDs, the development process of near-infrared (NIR) QDs is relatively slow because of the challenges, such as lacking of effective synthesis method and characterization techniques. Due to the above-mentioned advantages, NIR QDs show great potential and promise in many applications. Apart from the applications in optoelectronic devices, NIR QDs are also widely employed in interdisciplinary domains including their acting as optical amplifier medium for electrical communication systems,[12] as remote sensing or as fluorescent materials for biomedical imaging, labeling and sensing.[13-15] Among all types of the NIR QDs that had already been found,[6] the lead chalcogenide QDs (PbX; PbX=PbS, PbSe, PbTe) in IV-VI group have shown the great promise both in fundamental scientific research and in technological applications, as they have narrow bulk bandgaps (0.41 eV, 0.278 eV and 0.31 eV at room temperature, respectively) along with broadband absorption, large excitonic Bohr radii (due to the high dielectric constant and small effective mass of carriers) and multiple exciton generation.[16] Furthermore, their energy bandgap can be widely tuned from about 0.3 eV to >2.0 eV.[12-15, 17-20] These remarkable and unique properties make the PbX QDs ideal candidates in NIR optoelectronic applications.

In order to obtain the desired properties of colloidal NIR QDs, it is essential to opt for the suitable synthesis method and control the reaction conditions. For PbX QDs, the "hot injection method" synthesis involving the rapid nucleation and continuous growth is the most commonly used due to its low cost, simple preparation, and high production yield.[21-23] Synthesis of PbX QDs is similar to the method of cadmium chalcogenide QDs, and the precursors are typically similar. By controlling the reaction condition including the reactivity and concentration of precursors, the injection and growth temperature, and the growth time, a variety of crystal sizes and shapes can be achieved.[24] Therefore, understanding the hidden mechanism of nucleation and the subsequent growth are indispensable to attain colloidal NIR QDs with good quality. This mechanism is strongly concerned with the electronic structure of tunable molecular precursors, and most researchers focus more attention on the design of chalcogen molecular precursors than the metal ones.[25-27] Therefore, the advancement of rational choice of precursors plays a paramount role in the improvement of synthesis.

Another big stumbling block for NIR PbX QDs development is the stability issue. When exposed to the air in the ambient environment, PbX QDs will degrade quickly, leading to significant changes in electronic structure of QDs,[28] which obstructs the extensive application of PbX QDs and optoelectronic device fabrication. So far, plenty of efforts have been devoted to addressing this issue, including the design of inorganic or organic surface ligands,[29-32] and matrix engineering.[33-35] Apart from improving the stability of QDs, the ligands and matrix formed can also influence the charge transport of device. Significantly, forming a core/shell heterostructure by growing the extra inorganic shell is also an effective approach to efficiently stabilize the QDs.[36, 37] Hence, it is vital to understand the surface chemistry, which involves both the electronic coupling within QDs arrays[38-40] and the core QD/ligand properties.[41-44]

Thanks to the enormous efforts made by researchers, the intrinsic properties of PbX QDs have been improved a lot, which paves the way for PbX QDs-based optoelectronic devices fabrication. Particularly, the PbX QDs based light-emitting devices have received growing attention in the past several years. Device performance significantly depends on photoluminescence quantum yield (PL QY) as well as the charge transport efficiency.[45, 46] PL QY of PbX QDs can be improved by surface engineering during synthesis,[47, 48] among which the controllable synthesis of core/shell QDs[49, 50] and optimization of ligands are of vital importance.[51] In addition, adjusting the appropriate inter-dot spacing is an effective way to enhance the carriers mobility and charge balance efficiency using solid state ligand exchange[6, 52] method in device. Besides, incorporation of QDs into matrix has been demonstrated to efficiently reduce the self-quenching and thus to boost emission efficiency.[1, 53-55] Up-to-date, the best performance of NIR QDs based LEDs (NIR-QLEDs) of ~7.9% EQE[56] have been achieved and more remarkable progress will be obtained with the continuous efforts devoted in the future.

In this review, we illustrate the comprehensive progress of PbX based NIR-QLEDs (Fig.1). Firstly, we begin with the recent progress in the synthesis of colloidal NIR PbX QDs, followed by the design of tunable precursors. Then, we summarize the established strategies to control the size and to improve PL QY, stability, and scale-up of PbX QDs. Next, we focus on the latest advance in PbX-based NIR-QLEDs improved by the enhancement of carrier diffusion and the passivation of defects. Finally, we draw a conclusion and make an outlook on the future application of NIR QDs in light-emitting devices.

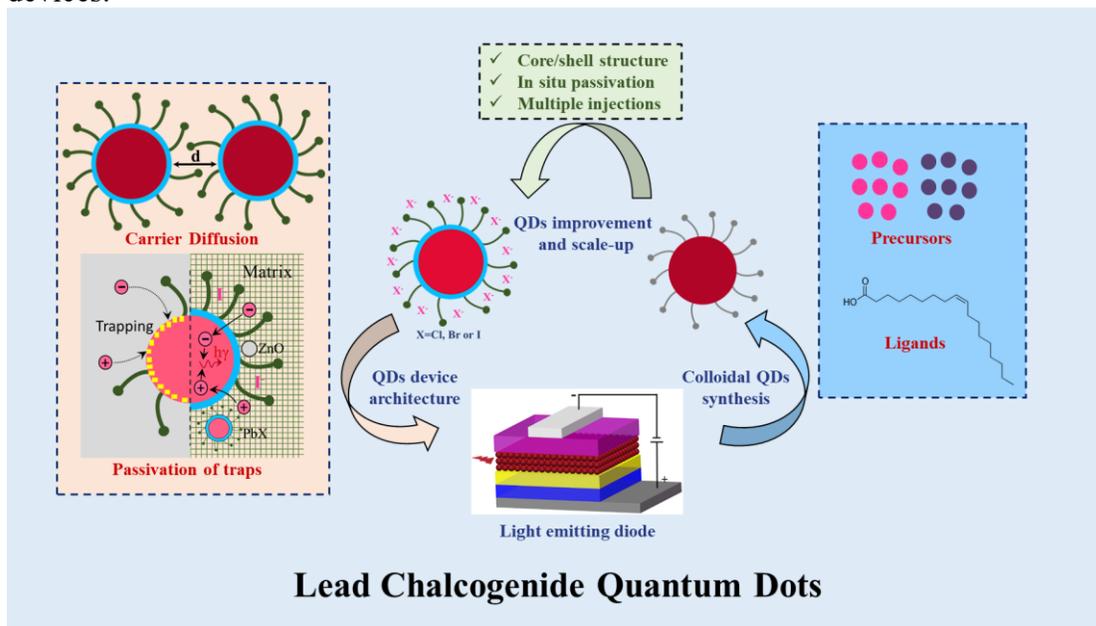

Fig.1. The comprehensive progress of PbX based NIR-QLEDs

## 2. Recent Advances in Lead Chalcogenide Quantum Dot Synthesis

单击此处输入文字。

### 2.1. Overview of Colloidal QD Synthesis

Typically, synthesis of QDs includes three main elements: precursors, ligands (or surfactants) and solvents. Precursors form highly active monomer during the reaction,[57] which are used to form small seed NCs and then grow larger with the increase of reaction time, and QDs with desired size will be obtained when the reaction is stopped. Ligands provide sufficient space or electrostatic repulsion to stabilize QDs and prevent uncontrolled aggregation of monomers and nuclei in the solvent, but QDs are still allowed to grow at high temperature. Therefore, the size and shape of QDs could be controlled by ligands. Furthermore, ligands could passivate the surface of QDs by eliminating harmful surface defects and suspended bonds of QDs. However, the type and the amount of ligands can unpredictably affect the chemical composition of QDs surface, optical and electrical properties, functionality and processability. In addition, the procedure of purification is significant for the quality of QDs.

Solvents also play important role. In the early days, most researchers synthesized colloidal QDs by traditional aqueous phase method, which has the deficiencies of low PL QY, wide particle size dispersions, and unwanted spectral tailing. Then the solvent for synthesis is gradually changed from water to organic solvents which have better effect. For example, because many organic solvent precursors have a higher boiling point than water, QDs with better crystal quality can be obtained by increasing experimental temperature.[58]

The PbX QDs formation process could be divided into three stages (Fig.2).[59, 60] In the first stage, the metal and chalcogen precursors react to form monomers of PbX QDs under specific conditions. In the second stage, the supersaturated monomers aggregate fast into some small nuclei until the monomers concentration is lower than the "nucleation critical concentration".  In the third stage, additional monomers will coat on the nuclei to grow into PbX QDs until the monomers concentration is depleted. At this time, the PbX QDs are in the dynamic equilibrium process of growth and dissolution. Ostwald ripening process).[61, 62] should not be overlooked at this stage that some smaller PbX QDs could dissolve into monomers and make some larger PbX QDs grow larger.

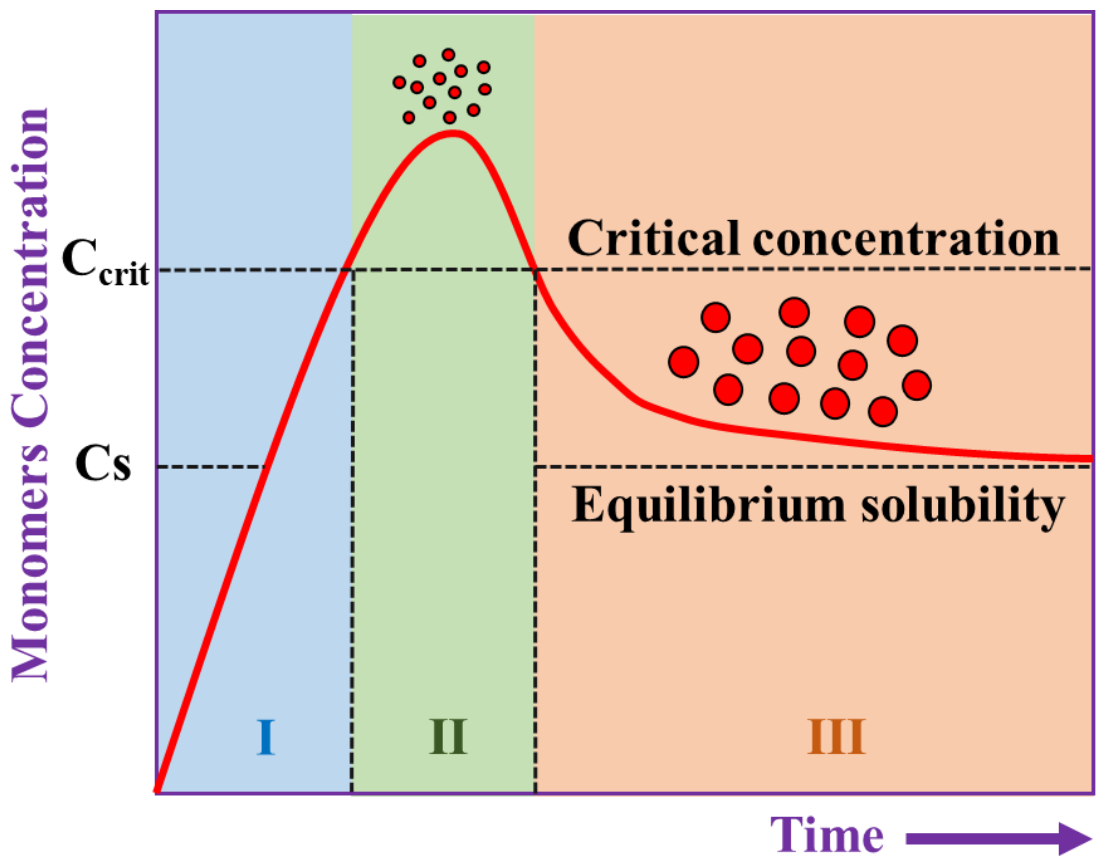

Fig.2. The PbX QDs formation process: prenucleation stage (I) nucleation (II) and growth (III) phases.

In the synthesis of colloidal PbX QDs, reaction temperature, precursor types and concentrations, precursor reaction types (e.g. exothermic or endothermic), stabilizers on the surface of QDs (ligands and surfactants), solvents and so on are important parameters, which can affect both size and morphology of PbX QDs. Therefore, rational control of experimental parameters is necessary for yielding high quality PbX QDs with desired size and morphology. For better utilizing PbX QDs in their applications, it is of critical importance to ensure the quality of PbX QDs. PL QY and stability in the ambient environment are two of the most vital technical indicators as mass production of PbX QDs with high quality for commercialization is attracting more and more attention.

## 2.2. Precursors for the Synthesis of Lead Chalcogenide Quantum Dots

PbX QDs are formed by the reaction of lead and chalcogenide precursors in one synthesis system with solvent and ligands. At present, the most commonly used solvents in the synthesis process are 1-octadecene (ODE) and oleic acid (OA).[63, 64] The OA is also the main choice for ligands of NIR QDs,[65, 66] which are used to stabilize QDs and passivate the dangling bonds on the surface of NIR QDs.

Lead oleates are one of the common lead precursors formed by reacting lead salt, e.g. $PbCl_2$,[67-69] PbO,[19, 70-73] and $Pb(CH_3COO)_2$[63, 64, 74-76] with OA[72, 73, 75] as the coordinating ligand in the non-coordinating solvents, such as ODE and diphenyl ether. In some case, OLA is used as the coordinating ligands. [68, 69]

There are many selenium sources which could be used to make selenium precursors in the synthesis of PbSe QDs, such as n-trioctylphosphine selenide (TOP-Se)[66, 77-79] and bis(trimethylsilyl)selenium (TMS-Se) ,[80, 81] SeO$_2$,[82] and selenourea.[83] In 2001, Murray *et al.* first successfully synthesized high-crystallinity PbSe QDs by hot injection of lead oleate and TOP-Se precursors, and the exciton absorption peak also extended from 1200 nm to 2200 nm.[65] Some researchers also synthesized a series of PbSe from 3 nm to 8 nm in diameter.[84] Over the next few years, others made appropriate changes based on the method of Murray, and successfully synthesized other types of NIR PbX QDs. In 2010, SeO$_2$ with high chemical stability was used as an selenium sources in the synthesis of PbSe QDs[82] conducted in air, without using air-free set up (e.g. glove box or Schlenk line) but SeO$_2$ material required a higher temperature for reaction, and the size of the synthesized QDs was also larger. Others like elemental selenium, selenourea in N, N-dimethylformamide (DMF), tris (diethylamino) phosphine selenide (TDP-Se) and oleylamine selenide (OLA-Se) are also used as selenium sources.[79] However, some precursors like selenourea needed to use size-selection precipitation to achieve narrow nanocrystals size distribution.[85] because it was not easy to control the rate of nucleation and growth process.[86]

Bis(trimethylsilyl)sulfide (TMS-S) [19, 71] was also commonly used as sulfur precursor in the synthesis of PbS QDs. In 2003, Hines *et al.* used a method of injecting TMS-S into a lead oleate precursor at a high temperature to obtain PbS QDs with a PL QY of 20-30% and a narrow size dispersion (10-15%).[19] There were also other materials could be used as sulfur sources for synthesizing PbS QDs, such as powdered sulfur,[68, 69] thioacetamide, sodium sulfide,[75] hydrogen sulfide gas, thioureas,[87] since many sulfur-containing compounds, the obtained PL QY was not very high and the tunable range was smaller than the QDs synthesized by TMS-S.

Trioctylphosphine telluride (TOP-Te) was usually used as tellurium precursor in the synthesis of PbTe QDs by dissolving powdered Te in TOP. In 2006, Murphy *et al.* synthesized PbTe QDs by injecting TOP-Te into the lead oleates that resulting from the reaction of PbO and OA in ODE.[18] In 2012, Pan *et al.* synthesized PbTe QDs by injecting TOP-Te into PbCl$_2$-OLA.[88] TeO$_2$ also was used as tellurium source by directly heating the mixture of TeO$_2$ and TOPO at elevated temperature, and Shen *et al.* used it to synthesize PbTe QDs in 2010.[89]

Since the TOP-Se used to synthesize PbSe QDs, the TMS-S used to synthesize PbS QDs, and the TOP-Te used to synthesize PbTe QDs are highly toxic and unstable in air, so many researchers focused on finding substitutable non-toxic precursors. Although other chalcogenide precursors mentioned above could also be used to synthesize high quality QDs, they still cannot completely replace the TOP-Se, TMS-S, and TOP-Te in synthesis due to either reaction requirement or QDs quality reasons.

## 2.3. Control for the Size of Lead Chalcogenide Quantum Dots

Because the wavelength of emission of PbX QDs is closely related to their sizes, there is no doubt that a well control of the PbX QDs size is of great importance for practical application. The number of synthesized QDs is largely dependent on the quantity of nuclei and the rate of nucleation, and the size of prepared QDs is also related to the rate of growth, which are both decided by the monomer supply kinetics. [87] It is well-known that the current PbX QDs synthesis is still very sensitive to reaction conditions, and thus it is necessary to control multiple parameters to obtain the desired QDs size, such as reaction temperature, reaction time, precursors and the surface capping ligands involved in the reaction.

In general, the higher the temperature allows fast nucleation, but also greatly promote the growth rate of PbX QDs, resulting in larger QDs at the same reaction time.[63, 71] For example, ultra-small PbSe QDs emitting visible light could be obtained at room temperature while mid-infrared emitting PbSe QDs could be achieved under high-temperature.[77] Typically, as the reaction time prolongs, the total number of QDs decreases because of Ostwald ripening, and the QDs size would gradually become larger. For PbSe QDs, the longer the reaction time is, the larger size of synthesized QDs can be achieved. On the contrary, the PbS NC synthesis is unique since the particle size is not strongly determined by the growth time due to very fast QDs formation process which ends within about several seconds.[71] Terminating the growth process of PbX QDs is an easy approach to control size, but with low chemical yield and batch to batch difference. In the synthesis system at full conversion, burst nucleation could conduct more uniform QDs nuclei,[90] which is of great importance of synthesizing monodisperse and smaller PbX QDs. Therefore, it is necessary to separate the nucleation from the growth process in the synthesis of high quality and monodisperse QDs.[91]

The monomer supply kinetics are closely related to the precursor reactivity and the monomer concentration. Fig.3 illustrates the formation of PbX QDs with high and low reactive precursors.[92] When the reactivity of precursors is high, the fast nucleation of a large number of monomers leads to the synthesis of particles with small size at a high concentration. Conversely, large particles with a lower concentration would be achieved by using low reactivity precursors. Controlled precursor reactivity and quantitative conversion could not only control the obtained PbX QDs size at industrially scales, but also improve the batch-to-batch consistency of PbX QDs.[87] In 2015, Zhang *et al.*[93] utilized Sn precursor to promote the nucleation of PbSe QDs, resulting in PbSe QDs with a size smaller than 3 nm. In 2015, Hendricks *et al.*[87] reported a tunable library of substituted thiourea precursors with different monomer supply kinetics thereby the extent of nucleation would be adjusted and finally PbX QDs with a desired size at full conversion could be obtained.

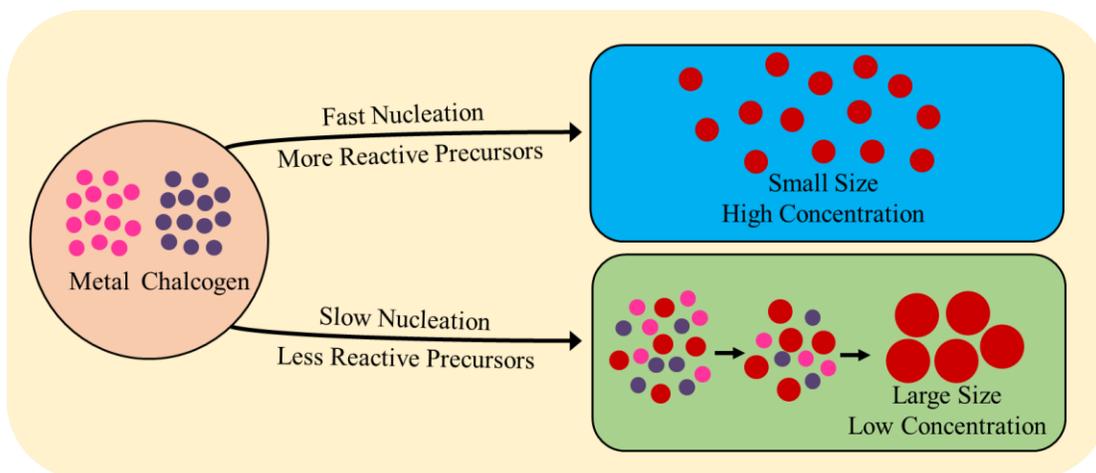

Fig.3. Illustration of colloidal nucleation and growth processes

Surface capping ligands of PbX QDs would also affect the QDs size. In 2007, Baek *et al.*[94] found that the sizes of PbSe QDs could be controlled by the chain length of organic ligands at a fixed reaction temperature and time. In 2009, Liu *et al.* [95] reported that the higher OA-to-Pb feed molar ratios made larger PbS QDs because the larger amount of OA would provide a better reaction medium solubility for the resulting PbS QDs. They also mentioned that the Pb-to-S feed molar ratio should be larger than 1 in case of the fast and out-of-control growth. In 2011, Yue *et al.*[75] reported that the average size of prepared PbS QDs decreased when increasing the concentration of OA, it was due to the reason that OA also influenced the reactivity of the supplied monomer and the higher concentration of OA would lead to a lower precursor reactivity, resulting in slower rate of QDs growth. In 2017, Ma *et al.*[96] introduced tributylphosphine (TBP) to synthesize ultrasmall PbS QDs (with diameter below 2.5 nm) because TBP slowed down growth process of PbS QDs.

## 2.4. Improvement of Photoluminescence Quantum Yield of Lead Chalcogenide Quantum Dots

The PL QY is the key factor to indicate the performance of PbX QDs. In the past few decades, the PL QY in core-only PbX QDs has been improved significantly.

PbSe QDs are usually synthesized by reaction of TOP-Se and lead precursor, and the improvement of PL QY in PbSe QDs is shown in Fig.4. In 2001, Murray *et al.*[65] synthesized the first PbSe QDs by rapidly injecting TOP-Se into lead oleate (lead acetate dissolved in OA), but PL QY of these PbSe QDs was not mentioned. In 2002, Krauss *et al.* [20] reproduced Murray's work, and the PL QY of PbSe QDs ranged from 12% to 81% with the emission wavelength ranged from 1200 to 1400 nm. The differences in the surface reconstruction during the reaction led to a wide range of changes in the PL QY. In 2004, Colvin *et al.*[70] synthesized PbSe QDs with very high PLQY of 85%. They dissolved PbO in OA and a noncoordinating solvent (ODE) as lead precursor instead of using lead acetate dissolved in OA, while the selenide precursor still used TOP-Se.

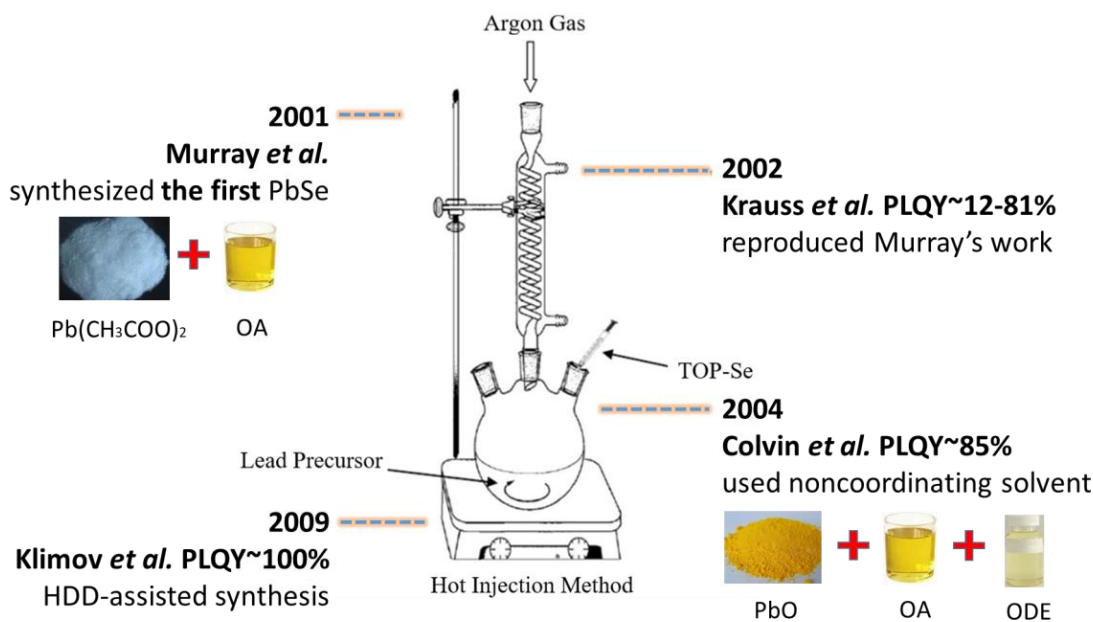

Fig.4. Improvement of PL QY in PbSe QDs. The data are taken from Ref.[20, 65, 70, 97]. Copyright 2002, American Chemical Society; Copyright 2004, American Chemical Society; Copyright 2009, American Chemical Society

In 2009, Klimov *et al.* [97] synthesized PbSe QDs with the highest PLQY of 100% using 1,2-hexadecanediol (HDD)-assisted synthesis. There were two possible mechanisms to form PbSe monomer,[98] under the effect of TOP, $Se^0$ and $Pb^{2+}$ obtained two electrons to become $Se^{2-}$ and $Pb^0$, then they reacted with $Pb^{2+}$ and $Se^0$ to form PbSe QDs, respectively, accompanied by the formation of Trioctylphosphine oxide (TOPO) and anhydride. The reducing agents could reduce TOPO to TOP, which was helpful to form more PbSe QDs. It was notable that the suitable reduce agent was important for improving chemical yield of PbSe QDs, and maintaining PL QY at the same time by well controlling the growth rates.

As shown in Fig.5, the PL QY of PbS QDs has increased from 20% to 90% from 2003 to 2011. In 2003, Hines *et al.* [19] synthesized the first PbS QDs using lead oleate (PbO dissolved in OA) and TMS-S diluted in ODE with the lead/sulfide molar ratio of 2:1. Their PbS QDs showed a good controllable wavelength from 800 to 1800 nm with a narrow size dispersion of 10-15%. And the PL QY of PbS QDs could reach 20%, which would decrease with the particle size increased.

In 2006, Cademartiri *et al.*[99] presented another route to synthesize PbS QDs with wavelength from 1245 to 1625 nm, and higher PL QY of 40% was achieved. The elemental sulfur (S) dissolved in oleylamine (OLA) was quickly injected into $PbCl_2$ dissolved in OLA. In this unique system, amine served as the capping group and $Cl^-$ ions balanced the charge on the surface of PbS QDs.

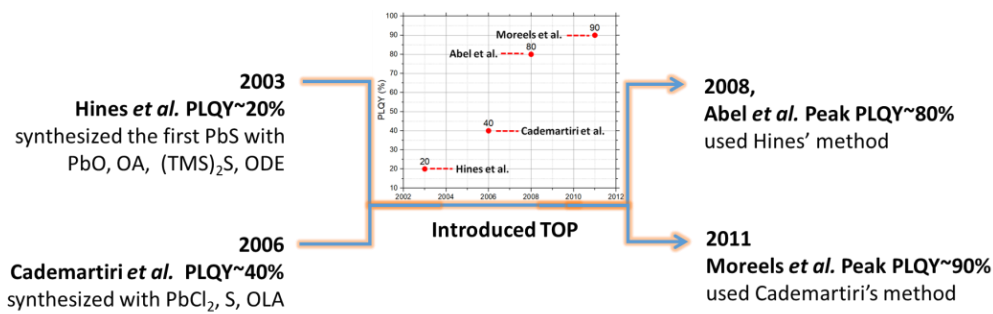

Fig.5. Improvement of PL QY of PbS QDs. The data are taken from Ref.[19, 99-101]. Copyright 2003 WILEY-VCH Verlag GmbH &Co. KGaA, Weinheim; Copyright 2006, American Chemical Society; Copyright 2008, American Chemical Society; Copyright 2011, American Chemical Society.

In 2008, Abel et al.[100] introduced TOP in the synthesis to improve the PL QY of PbS QDs based on Hines's synthesis, and the PL QY of 80% could be improved from 40%.

The TOP played an important role on the performance of the PbS QDs because OA ligands would well bind to lead atoms on the surface of PbS QDs, and sulfur atoms were not passivated and would trap hole, resulting in poor luminescence. The TOP introduced in the synthesis of PbS QDs could efficiently bond to the non-passivated sulfur atoms to decrease the hole defects, resulting in better luminescence. In 2011, Moreels et al.[101] introduced TOP in the synthesis of PbS QDs based on Cademartiri's synthesis, the PL QY could be improved from 40% to 90%.

## 2.5. Stability and Scale-up of Lead Chalcogenide Quantum Dots

High stability of QDs in the ambient environment will enable mass production and wide range of applications of PbX QDs in the future. PbX QDs are air-sensitive and light-sensitive, which are easy to be oxidized under the light and air. The destructive oxidization of PbX QDs usually causes the PL peak blue shifting and PL QY reduction. There are three approaches to improve stability of PbX QDs illustrated in this review including the core/shell structure, halide treatments, and multiple injections.

### 2.5.1. Core/Shell Structure

In order to enhance the stability of PbX QDs, a shell material with larger bandgap and better chemical stability is grown over the PbX QDs core to passivate trap states, which could confine the holes and the electrons in the core and protect the core from being exposed to air. As shown in Fig.6, there are two approaches to form core/shell structure including the cationic exchange and epitaxial shell growth. [102, 103] The coated shell may exist in three cases such as uniform shell, composition gradient shell or alloyed shell.

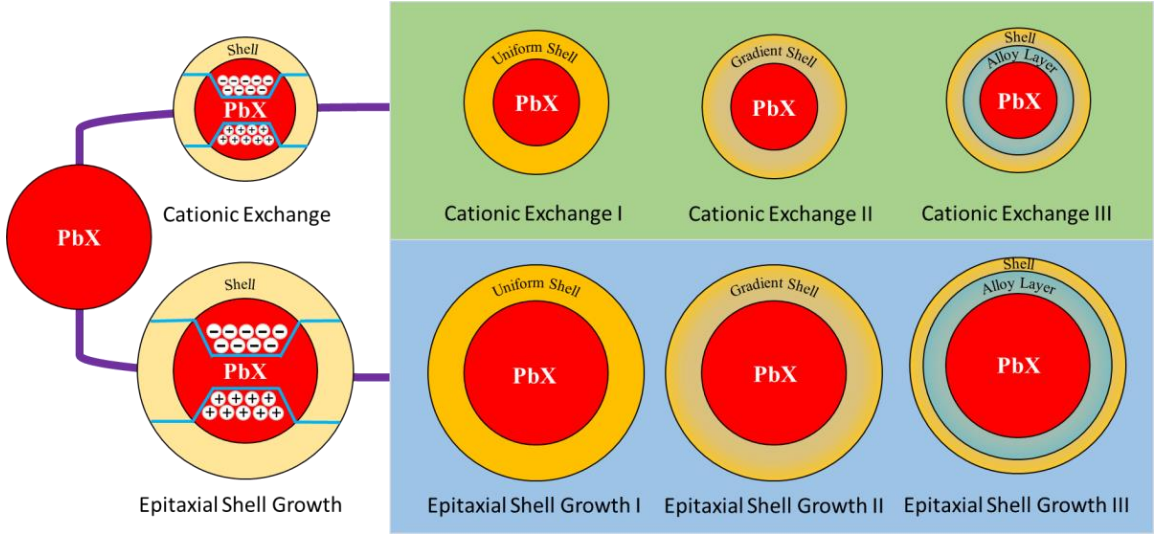

Fig.6. Core/shell structure of QDs after cation exchange and epitaxial shell growth.

Cationic exchange is the most common method to form the shell for PbX QDs because the Pb$^{2+}$ cations near the surface are susceptible to be exchanged with other metal ions in the solution such as Cd$^{2+}$ cations, which could be proved by the blue-shifted emission of resulted core/shell QDs compared to that of core-only QDs due to the reduced size of PbX QD. In 2008, Pietryga et al.[37] firstly produced PbSe/CdSe and PbS/CdS core/shell QDs using cationic exchange method, and those resulted QDs were stable against spectral attenuation and spectral shifting. Additional ZnS shell could also be coated on the surface of PbX/CdX core/shell QDs, which were likely to reduce toxic effects associated with exposure to heavy-metal ions. In 2010, Zegar et al.[104] synthesized PbSe/CdSe core/shell QDs by Pietryga's method. In 2011, Zhao et al. synthesized thick-shell PbS/CdS QDs with PL QY up to 67% and of much better photo- and thermal stability compared with those of core-only PbS QDs by two-step cationic exchange method. In the followed years, there were still many other groups[64, 105-107] reproduced PbX/CdX QDs based on Pietryga's method. And in 2013, Ren et al. reported a microwave-assisted cationic exchange method for the synthesis of PbS/CdS core/shell QDs through a uniform growth method, and those obtained QDs shown a PL QY of as high as 57% which was about 1.4 folds higher than that achieved by the same PbS/CdS core/shell QDs prepared using traditional heating in an oil bath. In 2017, Yang et al. made NIR-QLEDs based on high-quality PbS/CdS QDs followed by Ren's method.

Although cationic exchange method could improve PL QY and stability compared to core-only PbX QDs, it was still difficult to control of the interface between the core and the shell and fine tuning the thickness of the CdX shell. The Successive Ion Layer Adsorption and Reaction (SILAR) method proposed by Peng et al.[108] were used to coat epitaxial CdX shell on PbX QDs, increasing their stability over time which was attributed the protective CdX layer against oxidation. In 2010, Zhang et al.[109] obtained PbSe/CdSe core/shell QDs with high PL QY of 70% by SILAR method in solution, and a red-shifted emission could be observed with the growth of CdSe, which could be explained by the increased delocalization of carriers into the shell[102]. They proved that the SILAR method could exactly controlled the thickness of CdSe shell and largely improved the stability of PbSe

QDs, however, the SILAR method was also performed at temperature as high as that could be used in cationic method, which might lead to an ion exchange reaction, and it also needed to carefully avoid secondary nucleation of the excess shell precursors. In 2016, Sagar *et al.* demonstrated a colloidal atomic layer deposition growth (c-ALD) method for the synthesis of PbS/CdS core/shell QDs at room temperature, which largely preserved the pronounced spectral features of PbS QDs and fine control over the CdS shell growth. But they did not demonstrate PL QY enhancement by the epitaxial CdS shell growth on the PbS QDs, which might be caused by the nonradiative recombination by defects at the interface between PbS and CdS. In 2017, Bawendi *et al.* also used room-temperature c-ALD method to synthesize PbS/CdS core/shell QDs, and they found the PL QY and the stability against photobleaching of resulted PbS/CdS QDs could be further tremendously improved by the Cd(OA)$_2$ treatment.

There were also some group combined cationic exchange and SILAR together. In 2016, Zhao *et al.* synthesized PbS/CdS/CdS core/shell/shell QDs with ultra large shells by using SILAR method to deposit CdS layer on the prepared PbS/CdS QDs by two-step cationic exchange method. In 2017, Hollingsworth *et al.* also sequentially applied partial-cation exchange method and SILAR method to reduce giant PbSe/CdSe/CdSe core/shell/shell QDs, which were highly stable for both photobleaching and blinking were suppressed.

### 2.5.2. Halide Treatments
Halide treatments are attractive to improve the stability of PbX QDs because the resulting ligands are compact to avoid oxidation and the halide ions could balance excess charge resulting from nonstoichiometric surface termination, which could be realized by using halide precursors before the reaction or injecting halide salts during the reaction. In 2013, Zhang *et al.* [30] introduced the halide in the synthesis of PbS and PbSe QDs. Sulfur precursor or selenide precursor was injected into lead precursor. PbY$_2$ (Y = Cl, Br, I) served as both the lead source and the capping ligand, as shown in Fig.7, the halide atoms bonded to the surface of PbX QDs, which improved the photostability of both PbS and PbSe QDs stored under air. In 2014, Ju Young Woo *et al.* [110] achieved air-stable PbSe QDs, by adding halide salts (PbCl$_2$ or PbI$_2$) after synthesizing PbSe QDs. In 2015, Yi Pan *et al.* [111] used Zhang's method [30] to synthesize PbSe QDs and found that the PbSe QDs remained stable in air for 24 months.

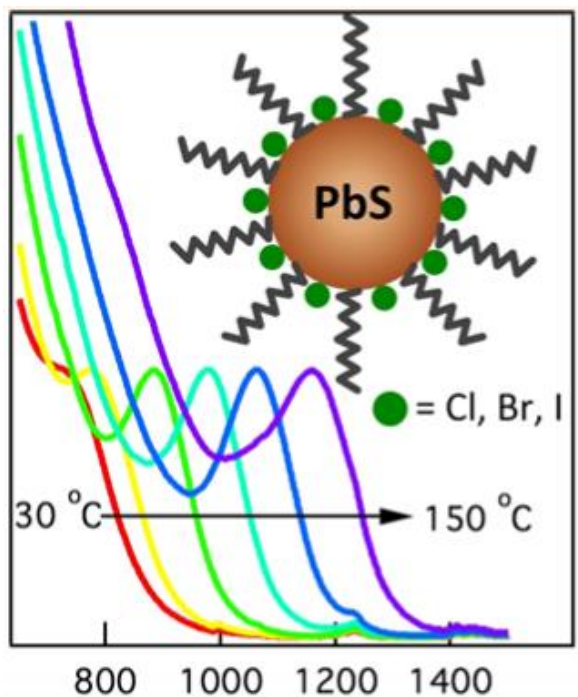

Fig.7. PbS QDs with halide treatments. Reproduced with permission from Ref.[30]. Copyright 2011, American Chemical Society.

### 2.5.3. Multiple Injections

In 2016, Franky *et al.*[112] synthesized PbS QDs based on Weidman's synthesis, which used multiple injections of the sulfur precursor to grow the PbS QDs without initiating new nucleation during growth and Ostwald ripening.[67] These QDs shown great stability and the resulting devices were stable even without encapsulation.

However, the PbX QDs could not be upscale because of the long-time synthesis process, the complicated operation, and high cost. Adjust the temperature and reaction time were the simplest operation of synthesis. In 2006, Cademartiri *et al.* [99] synthesized 1.5 g PbS QDs from a 50 mL reaction by a solventless technique. In 2013, Zhang *et al.*[30] batched up PbS QDs based on Hines and Cademartiri approach, and PbS QDs of 47 g were achieved. It was also a challenge to add large volumes of the precursors in current hot-injection protocols for mass production. In 2017, Yarema *et al.*[113] reported an under-pressure-assisted method to upscale QDs for the large-batch synthesis (Fig.8), which allowed fast addition of large injection volumes, and 17.2 g PbS QDs were finally achieved. In addition, those creative additives (such as the secondary phosphine sulfide precursors[72, 114, 115] or the thiourea precursors[87]) could not only improve nucleation of QDs, but also upscale the QD quantity.

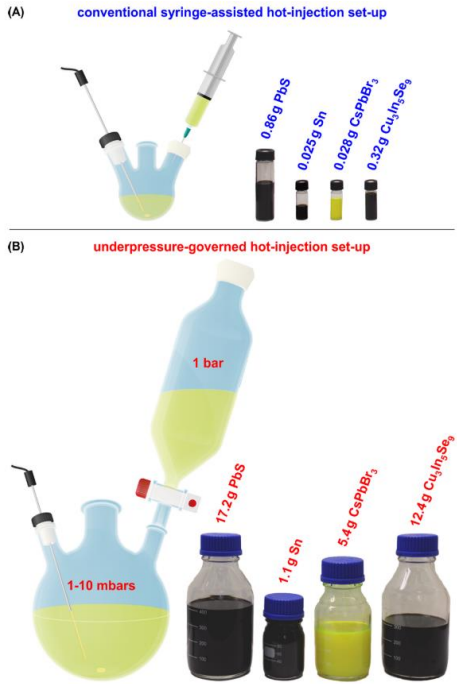

Fig.8. Schematic illustration of setups and obtained products for (a) conventional syringe-assisted hot-injection and (b) underpressure-governed hot-injection methods.[113]

## 3. Recent Advance in Lead Chalcogenide Quantum Dot based Light Emitting Diodes

The NIR LEDs are currently fabricated by epitaxial growth of direct bandgap semiconductors, which makes them expensive and difficult to integrate with other materials. However, colloidal semiconductor QDs can be produced by low-cost solution-process and can be directly integrated with silicon, which makes them become potential candidates for NIR LEDs in the future. The PbX QDs is suitable to be used in NIR-QLEDs, whose emission wavelengths can be tuned by tuning the size of QDs.

In this section, the recent progress of NIR-QLEDs is reviewed. The developing roadmap of PbX based NIR-QLEDs over recent years is shown in Fig.9. The first PbX based NIR-QLEDs was reported in 2000 with external quantum efficiency (EQE) of 0.5%, and there is a dramatic improvement in the EQE of NIR-QLEDs, from 2% in 2012 to 7.9% in 2018.

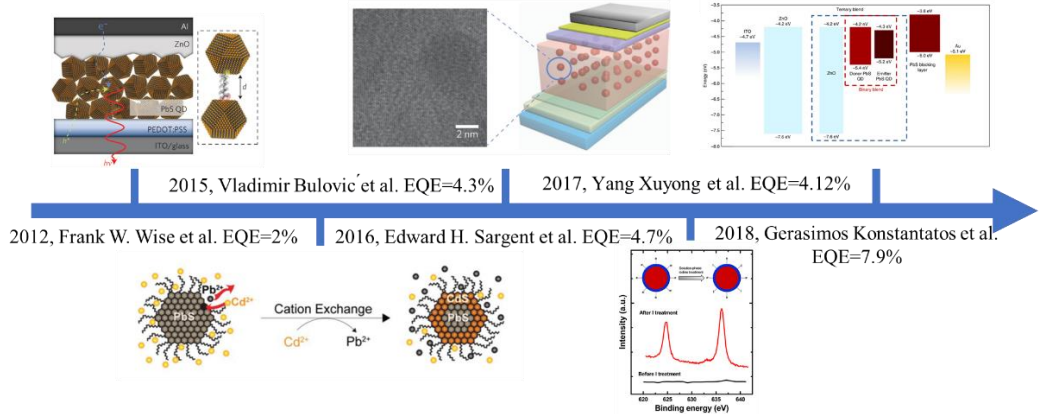

Fig.9. Development of PbX QDs based NIR-QLEDs over recent few years. Reproduced

with permission from Ref.[6, 7, 33, 56, 107]. Copyright 2012, Springer Nature; Copyright 2015 WILEY-VCH Verlag GmbH & Co. KGaA, Weinheim; Copyright 2016, Springer Nature; Copyright 2017, Springer Nature; Copyright 2018, Springer Nature.

As an important parameter to evaluate the efficiency of NIR-QLEDs, the EQE, can be expressed by the following equation[33]:

$$EQE = \eta_{diff}\, \eta_{inj}\, \eta_{PLQY}\, \chi\, \eta_{out}$$

where $\eta_{diff}$ is the efficiency of injected carriers that successfully diffuse to QDs; $\eta_{inj}$ is the efficiency of these carriers that transfer into QDs and form excitons; $\chi$ is the efficiency of these excitons whose states have spin-allowed optical transitions (for colloidal QDs, $\chi =1$), $\eta_{PLQY}$ is the internal QDs PL QY, $\eta_{out}$ is efficiency of light coupling out. Therefore, it is clear that $\eta_{diff}$ and $\eta_{PLQY}$ are two of key parameters for the efficiency of NIR-QLEDs device. How the performance of NIR-QLEDs devices is enhanced by improving carriers diffusion in the devices and PL QY of QDs will be discussed in section 3.1 and section 3.2, respectively. In visible-QLEDs, it is significant to choose suitable carrier transport layer materials[116-118] to improve $\eta_{inj}$. However, there is no related research specifically on this issue in NIR-QLEDs, which may become a promising research direction in this field. In our review, one method ("inter-dot spacing controlling") to improve carrier diffusion efficiency and four methods ("core/shell QDs", "perovskite matrix", "iodide capped QDs", "ternary blend of QDs film") to increase PL QY will be discussed as shown in Fig.10.

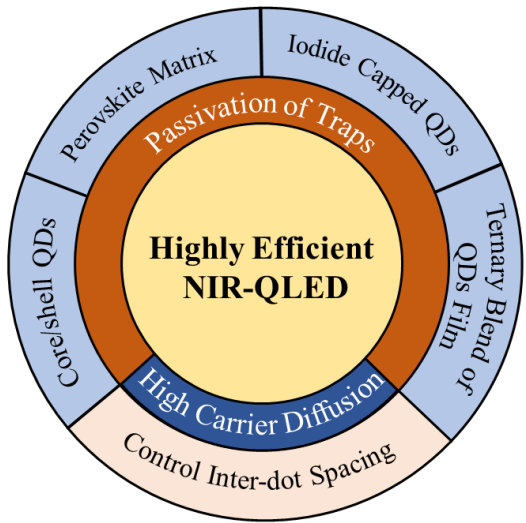

Fig.10. Main strategies and specific methodologies for highly efficient NIR-QLEDs.

### 3.1. Improvement of Carrier Diffusion in Lead Chalcogenide Quantum Dot based LEDs

Long alkyl chains insulating molecules, such as OA and OLA, are usually used as ligands in the synthesis of PbX-based QDs. However, those insulating ligands will greatly affect the performance of NIR-QLEDs caused by blocking inter-dot carrier transfer. Therefore, it is necessary to replace the original long alkyl chains with shorter ligands to enhance the diffusion of carriers for high-efficiency NIR-QLEDs. In 2012, Frank W. Wise *et al.*[6] in Cornell University studied the effects of different length ligands on PbS-based NIR-QLEDs. They used ligands with different carbon chain lengths (such as MPA, MHA, MOA, MUA) to control the inter-dot spacing in the emitting layer by solid-state ligand exchange. They found that the inter-dot spacing became smaller as the length of the surface ligands decreased. On the one hand, the diffusion of carriers in the emission layer was increased so that increasing the exciton recombination; on the other hand, too short inter-dot spacing would enhance the probability of exciton separation and reduce the exciton recombination. In addition, surface ligands of different lengths also affected the light extraction efficiency and carrier balance of NIR-QLEDs. Their experimental results (Fig.11(a)) showed that the 8-carbon chain (8-mercaptooctanoic acid, MOA) ligand was the best choice for NIR-QLEDs (energy-level diagram of device as shown in Fig. 11(b), which first achieved 2% peak EQE and 6.4 W Sr$^{-1}$ m$^{-2}$ radiance at 1054 nm.

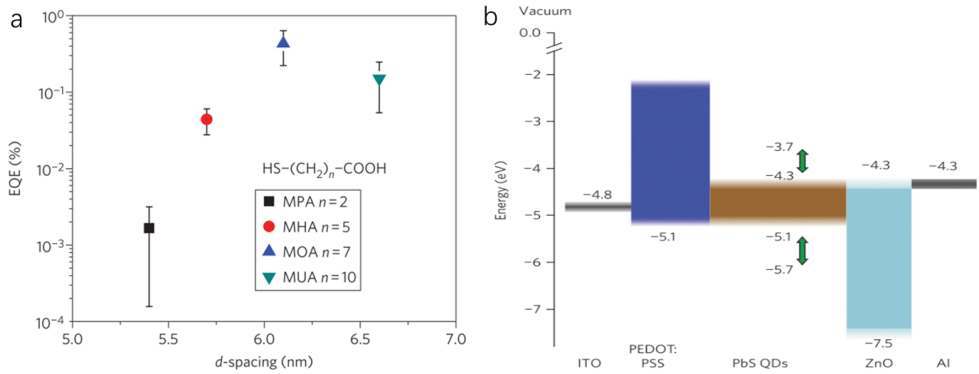

Fig.11. (a) EQE dependence on inter-dot spacing; (b) Energy-level diagram of the device. Reproduced with permission from Ref.[6]. Copyright 2012, Springer Nature.

### 3.2. Passivation of Traps in Lead Chalcogenide Quantum Dot based LEDs

Low PL QY of NIR QDs films was another reason for the poor EQE of NIR QLEDs, which was mainly caused by self-quenching, wherein inter-dot transport facilitated transport-assisted trapping, leading to non-radiative recombination even at relatively rare defect sites. Four methods ("core/shell QDs", "perovskite matrix", "iodide capped QDs", "ternary blend of QDs film") were used to passivate NIR QDs film to improve device performance in recent years (as shown in Fig.12).

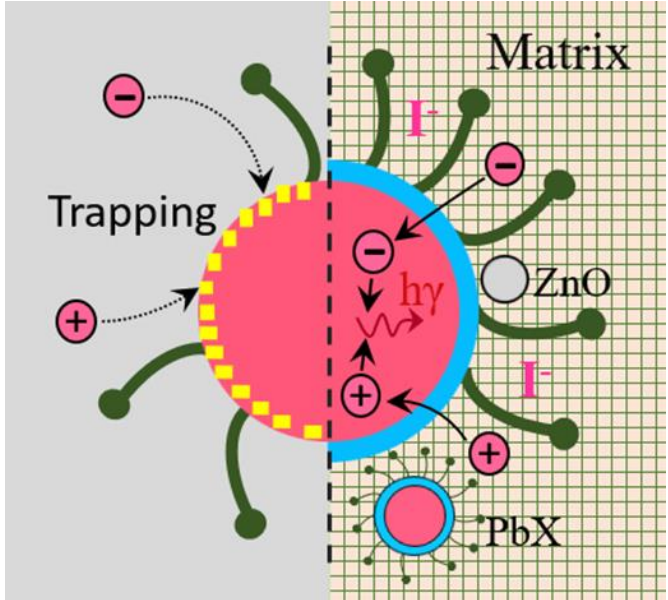

Fig.12. Passivation of traps in PbX based NIR-QLEDs.

In 2015, Vladimir Bulovic *et al.*[107] in Massachusetts Institute of Technology improved the performance of NIR-QLEDs by replacing the PbS QDs to PbS/CdS core/shell QDs by cationic exchange. On one hand, the CdS shell could significantly passivated PbS QDs against in situ nonradiative pathways in NIR-QLEDs (as shown in Fig.13(a)). On the other hand, most of the carrier would be confined in the core because the band gap of the shell was larger than the core's, and the CdS shell could protect PbS QDs from the damage of water and oxygen. Therefore, the PL QY and stability of the QDs were both significantly improved. The PL QY of PbS based QDs films could be increased by 100 times, and efficient PbS/CdS QDs based NIR-QLEDs of 4.3±0.3% EQE at 1242 nm were achieved (energy-level diagram of device as shown in Fig.13(b).

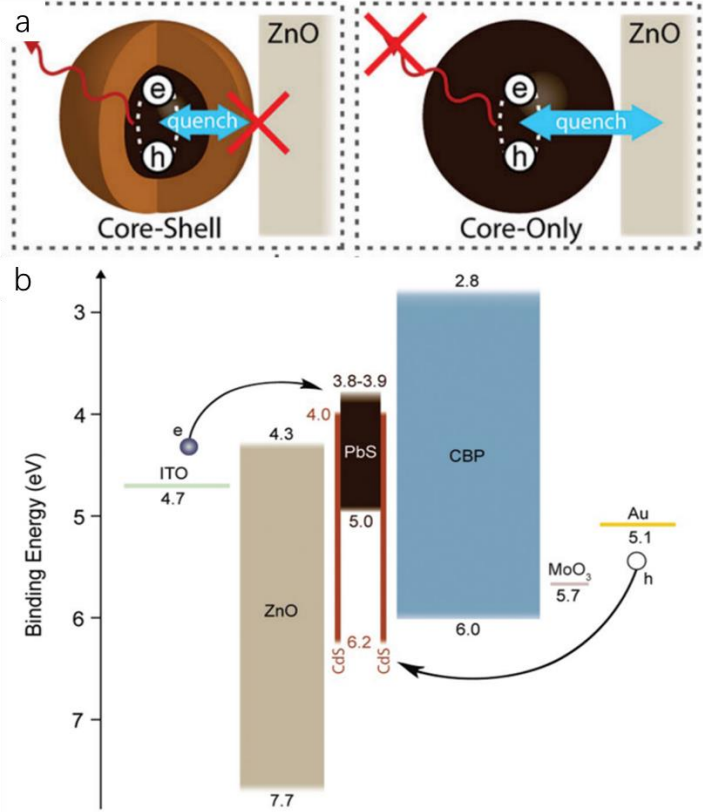

Fig.13. The CdS shell could significantly passivated PbS QDs against in situ nonradiative pathways in NIR-QLEDs. Reproduced with permission from Ref.[107]. Copyright 2015 WILEY-VCH Verlag GmbH & Co. KGaA, Weinheim.

In 2017, based on the work mentioned-above, Xuyong Yang *et al.*[7] in Shanghai University improved PL QY of core/shell PbS/CdS QDs film by iodine treatment in order to further improve the performance of NIR-QLEDs. As shown in Fig.14(a), The PL emission intensity of the I⁻ ligand capped PbS/CdS QDs film was 1.56 times higher than that of untreated PbS/CdS QDs film because the capped I⁻ could passivated the surface defects of PbS/CdS QDs. In addition, the capped I⁻ on PbS/CdS QDs played a significant positive role on the charge transport process that the current injection efficiency of the NIR-QLEDs was higher than that of without I⁻ treated based NIR-QLEDs. These significant improvement in the surface characteristics of PbS/CdS QDs contributed to the performance enhancement of NIR-QLEDs, and a maximum EQE of 4.12% and a maximum radiance of 6.04 W Sr⁻¹ m⁻² at 1510 nm was achieved (energy-level diagram of device as shown in Fig.14(b).

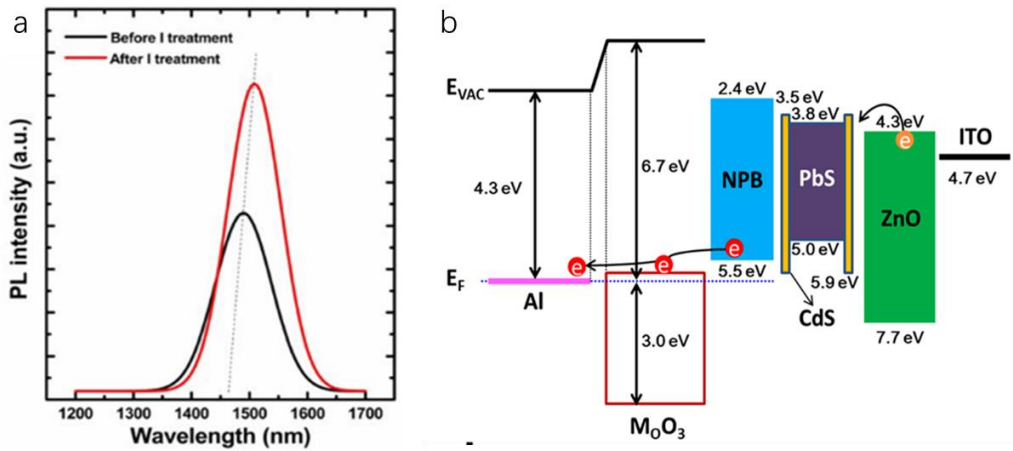

Fig.14. (a) The PL intensity of solution-phase iodine treated QDs films vs. the untreated QDs films; (b) Energy-level diagram of the device. Reproduced with permission from Ref.[7]. Copyright 2017, Springer Nature.

In 2016, Edward H. Sargent *et al.*[33] in University of Toronto selected perovskite as matrix of PbS QDs to improve the performance of NIR-QLEDs because perovskite had higher carrier mobility and fewer defect states than other matrix materials such as polymers and inorganic crystals. On one hand, the perovskite matrix enabled efficient transportation photo-generated carriers to PbS QDs, and the diffusion length of carriers could be largely enhanced. On the other hand, $MAPbI_xBr_{3-x}$ perovskite matrix could naturally passivate the surface of PbS QDs, and the strain at the interface between perovskite and PbS QDs and the density of defects were reduced because the lattices of PbS QDs and $MAPbI_xBr_{3-x}$ perovskite matrix matched well. These reasons mentioned above improved the PLQY of the QDs. As shown in Fig.15(a), the PL QY of QDs was enhanced and reached the peak at 50% iodine content when increasing the bromine concentration. And efficient NIR-QLEDs of $4.7 \pm 0.4\%$ EQE at 1391 nm were finally achieved (device structure as shown in Fig.15 (b).

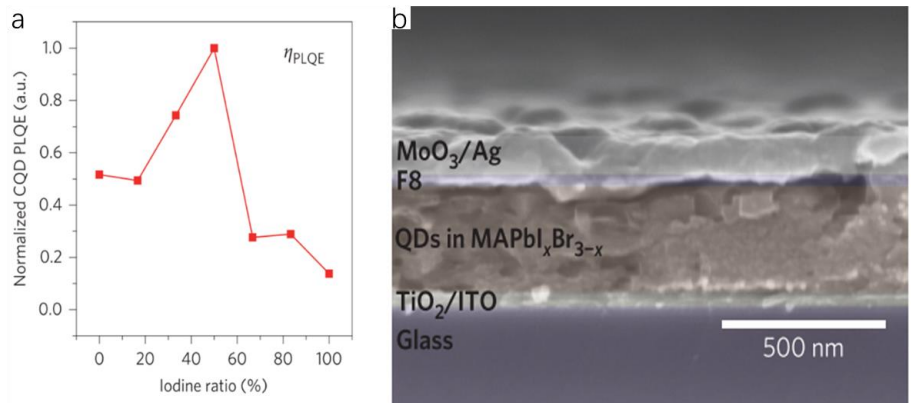

Fig.15. (a) PL QY of PbS QDs in $MAPbI_xBr_{3-x}$; (b) SEM image of device. Reproduced with permission from Ref.[33]. Copyright 2016, Springer Nature.

In 2018, Gerasimos Konstantatos *et al.*[56] in Institució Catalana de Recercai Estudis Avançats (ICREA) fabricated high-efficiency NIR-QLEDs by blending ZnO nanoparticles (NPs) and two different sizes of PbS QDs. The defects in PbS QDs film were passivated and the carrier supply to the emitting layer could be balanced by the ternary blend method. As shown in Fig.16(a), PL QY of PbS QDs films was increased by introducing smaller PbS QDs matrix into defective PbS QDs and reached the peak of 60% while the PL QY of single PbS QDs films was only 2.3%. It was explained that the small PbS QDs matrix could absorb excited light and possessed long carrier diffusion lengths to supply the emitting QD species with carriers, therefore, the number of proportionally defective large PbS QDs existence in the same volume was lower. The n-type ZnO NPs would passivate electron traps and enhance radiative recombination in the PbS QDs film. The PL QY of ternary blend films was increased to 80 % at 20% ZnO NPs content of the emitting layer. Mix ligands exchange (MPA and $ZnI_2$) was use to further improve the efficiency of NIR-QLEDs because mixed ligands could improve carrier's mobility and passivate defects, and mixed ligand exchange based NIR-QLEDs showed lower turn-on voltage and higher radiance than that of single MPA ligand. Finally, efficient NIR-QLEDs of 7.9% EQE and Radiance of 9 W $m^{-2}$ $Sr^{-1}$ at 1400 nm were achieved, which was the highest performance of NIR-QLEDs up-to-date (energy-level diagram of device as shown in Fig.16(b).

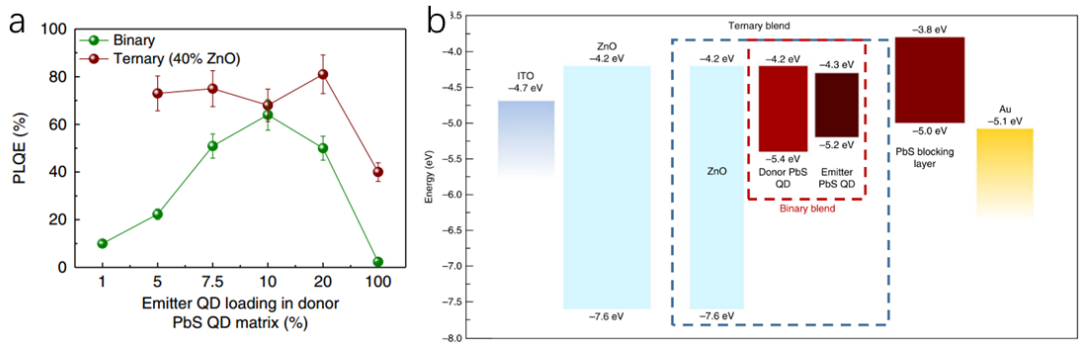

Fig.16. (a) PL QY of binary and ternary QD blends with different emitter QD loadings in the donor QD matrix; (b) Energy-level diagram of the device. Reproduced with permission from Ref.[56]. Copyright 2018, Springer Nature.

## 4. Conclusions and Perspectives

High quality colloidal PbX (PbSe, PbS, and PbTe) QDs can pave the way for the bright, low-cost, and high-efficiency NIR-QLEDs application, which also can be incorporated into flexible substrates. In this review, we summarized some of the recent achievements about synthesizing colloidal PbX-based QDs and important advances in the application of PbX QDs based NIR-QLEDs. Based on our critical analysis we conclude:

a) Over the past decade, a lot of progress has been achieved to improve the quality of colloidal PbX QDs based on the synthesis techniques of cadmium QDs. Although current methods could synthesize lab-scale NIR PbX QDs, it is still hard to realize their commercialization. New and suitable precursors are still needed to be explored, and the further research about the fundamental mechanisms of the PbX QDs nucleation and followed growth are needed to obtain desired PbX QDs with higher performance at lower cost. Besides, the intensive study of the effect of ligands on the surface of these QDs and the influence of protected matrix around them is also important for the sake of promoting the commercialization of colloidal NIR PbX-based QDs. Finally, innovative works about PbTe-based QDs are expected because the research progress in them was much slower than that of PbS-based and PbSe-based QDs until now.

b) Over recent years, the performance of PbX QDs based NIR-QLEDs has been improved largely, but relevant research work is still relatively few and the efficiency is still lower than visible-QLEDs. More works need to be focused on the improvement of NIR emitting layer, such as exploring new ligands or matrix for PbX-based QDs in order to further enhance the carrier diffusion, passivate the harmful traps of QDs, protect them from water and oxygen in the air, and improve the stability of light and heat during working. Moreover, the experiences of visible-QLEDs could be used for reference to enhance the EQE of NIR-QLEDs, such as balancing the injection of electrons and holes, enhancing light extraction efficiency, etc. Ongoing progresses of PbX QDs based NIR-QLEDs are expected to bring unprecedented discoveries soon.

**Acknowledgment**

The authors would like to thank the Pico Center at SUSTech that receives support from Presidential fund and Development and Reform Commission of Shenzhen Municipality.